\documentclass[twocolumn,showpacs,preprintnumbers,amsmath,amssymb,prl]{revtex4}

\usepackage{amsmath}
\usepackage{graphicx}
\usepackage{graphicx}
\begin{document}
\title{
Multimodal transition and stochastic antiresonance
in squid giant axons
}

\author{L.~S.~Borkowski}

\affiliation{Faculty of Physics, Adam Mickiewicz University,
Umultowska 85, 61-614 Poznan, Poland}


\begin{abstract}
The experimental data of
N. Takahashi, Y. Hanyu, T. Musha, R. Kubo, and G. Matsumoto,
Physica D \textbf{43}, 318 (1990),
on the response of
squid giant axons stimulated by periodic sequence
of short current pulses is interpreted
within the Hodgkin-Huxley model.
The minimum of the firing rate as a function of
the stimulus amplitude $I_0$
in the high-frequency regime is due to the multimodal
transition.
Below this singular point only odd multiples
of the driving period remain and the system
is highly sensitive to noise.
The coefficient of variation has a maximum
and the firing rate has a minimum
as a function of the noise intensity which
is an indication of the stochastic coherence antiresonance.
The model calculations
reproduce the frequency of occurrence
of the most common modes in the vicinity of the
transition.
A linear relation of output frequency vs. $I_0$ 
for above the transition is also confirmed.
\end{abstract}
\pacs{87.19.ll,87.19.ln,87.19.lc}
\maketitle


The Hodgkin-Huxley (HH) model\cite{HH1952}
is a prototypical resonant neuron
with the main resonant frequency typically
of order 40 to 60 Hz.
Its output interspike intervals (ISI) can be classified
in terms of integer multiples of the driving period.
The multimodality is revealed when the HH neuron
is stimulated by noisy inputs, such as additive noise\cite{Tiesinga2000,Arcas2003}, random synaptic inputs\cite{Brown1999,Tiesinga2000,Luccioli2006}
or channel noise\cite{Rowat2007}.
Such ISI histograms
are encountered frequently in periodically forced
sensory neurons.
An explanation in terms of a two-state system with
noise was put forward by Longtin et al.\cite{Longtin1991}.
The multimodal character is manifest
also in a deterministic HH model near excitation
threshold\cite{Clay2003,Borkowski2010}
and in regimes of irregular response between mode-locked states\cite{Borkowski2010}.
It was shown recently that also the parity of ISI
plays a significant role\cite{Borkowski2009}.
Even (odd) modes dominate in the vicinity
of even (odd) mode-locked states, respectively.
The most significant manifestation of this effect
is the multimodal odd-all transition between
states 3:1 and  2:1\cite{Borkowski2009}, where
the coefficient of variation (CV) has a maximum
and the firing rate has a minimum.
The notation \textit{p:q} means
\textit{p} output spikes for every \textit{q} input current pulses.
Below this singularity only odd multiples
of the input period exist and above it
harmonics of both parities participate in the response.
The transition may be crossed by varying either
the stimulus amplitude or the input period.
The minimum of the firing rate occurs slightly above
the transition.

In earlier experiments in giant axons of squid
stimulated periodically by a train of short rectangular current
pulses the firing rate, defined as the ratio
of the output and input frequency $f_o/f_i$,
had a well pronounced minimum
as a function of the interval between
adjacent pulses\cite{Matsumoto1987}
or the stimulus amplitude\cite{Takahashi1990}.
Even modes were absent below the minimum\cite{Takahashi1990}.
This effect occurred near the excitation threshold,
between states 3:1 and 2:1.
Another interesting result
was the continuous relation between the firing rate
and the stimulus amplitude.
This set of experimental and theoretical results
deserves a more detailed comparison.

The theory can be tested also by considering
a periodic drive in the presence of noise.
Noisy biological systems\cite{Wiesenfeld1995,Doiron2000,Stacey2001,Rudolph2001,Tiesinga2000,Luccioli2006},
including the HH neuron,
are known to exhibit stochastic resonance (SR).
This phenomenon is mainly, though not exclusively,
characterized by a maximum of the signal to noise
ratio as a function of the noise intensity.
Another effect associated with the presence
of noise is the decrease of the firing threshold
and the coherence resonance\cite{Gang1993,Pikovsky1997},
where the minimum variability of the output signal, expressed
by CV in absence of a deterministic drive,
is achieved at some intermediate noise strength.
Recently it was found experimentally\cite{Paydarfar2006,Sim2007}
theoretically\cite{Gutkin2008,Gutkin2009}
that small amplitude noise may decrease the firing rate
or even turn it off.
The nonlinear system in the vicinity of the multimodal
transition is a natural candidate for finding
interesting effects due to noise
since the trajectories of different modes
are very close in parameter space.
In the following we compare experimental data
to theoretical results for the deterministic
case and calculate the sytem's response
to a periodic drive with additive Gaussian noise.

\begin{figure}[t]
\includegraphics[width=0.39\textwidth]{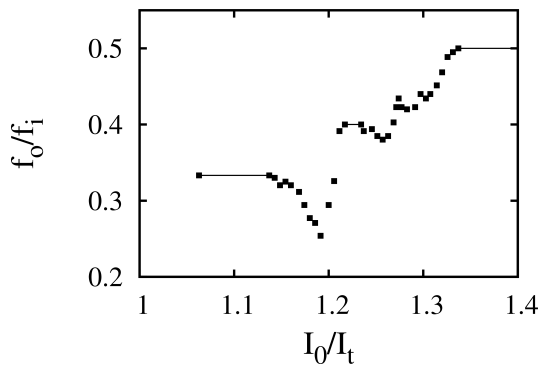}
\caption{The average firing rate, $\bar{T_o}/T_i$,
as a function of the stimulus amplitude $I$ from the work
of Takahashi et al.\cite{Takahashi1990}. $I_t$ is the minimum current
threshold obtained in the range $T_i=2.5 \textrm{ms}$ to
$T_i=6.5 \textrm{ms}$.
The measurements were carried out at $T_i=3.8 \textrm{ms}$.
}
\label{taka12}
\end{figure}

In the experiment of Takahashi et al.\cite{Takahashi1990}
the squid axon was stimulated by periodic train
of rectangular current steps of width $0.6 \textrm{ms}$.
Fig. \ref{taka12} shows the experimentally obtained
firing rate as a function of stimulus
amplitude scaled by the minimum
current threshold $I_t$.
On the left side of the minimum
only odd modes were recorded.
Even modes were present at the minimum point, with the $6:1$
mode occurring more frequently than the $4:1$ component,
and $2:1$ entirely absent.
This is consistent with calculation results\cite{Borkowski2009},
where even modes disappear
before reaching the multimodal transition (which is slightly
below the minimum of the firing rate), with the low
order modes vanishing first, beginning with mode $2:1$.

\begin{figure}[th]
\includegraphics[width=0.39\textwidth]{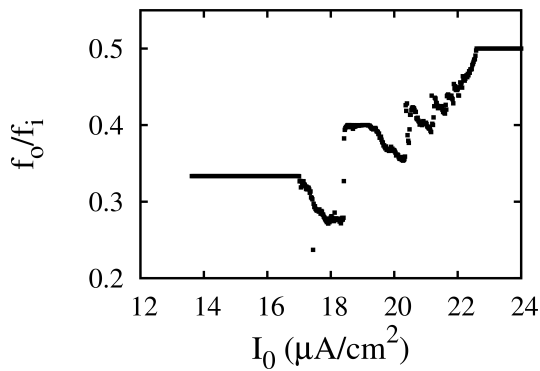}
\caption{The calculated average firing rate
at $T_i=7 \textrm{ms}$ without noise.
}
\label{taka12-fit}
\end{figure}

We try to reproduce this type of dependence using the HH model
with the classic parameter set and rate constants\cite{HH1952},
\begin{equation}
\label{HH}
C\frac{dV}{dt}=-I_{Na}-I_K-I_L+I_{app},
\end{equation}
where $I_{Na}$, $I_K$, $I_L$, $I_{app}$,
are the sodium, potassium,
leak, and external current, respectively.
$C=1\mu\textrm{F/cm}^2$ is the membrane capacitance.
The input current is a periodic set
of rectangular steps of width $0.6 \textrm{ms}$ and height $I_0$.
Equations are integrated within the fourth order Runge-Kutta scheme
with a time step of $0.001 \textrm{ms}$. The data points
are obtained from runs of 400 s, discarding the initial 4 s.
The dependence of the firing rate on the stimulus amplitude
is shown in Fig. \ref{taka12-fit}, where
$T_i=7 \textrm{ms}$.
The similarity to experiment
is striking. Although the calculated minimum
occurs at almost twice the experimental $T_i$,
the other time scales such as the refractory period
and the time span of the bifurcation diagram
differ by a similar factor. The entire dynamics
of the axon from the study of Takahashi et al.
is significantly faster than that of Hodgkin and Huxley.
This difference of time scales is not unusual.
Long time ago Best\cite{Best1979} noted that
the axon used by Hodgkin and Huxley was of poor quality
and in later studies significantly higher conductivities
were obtained.
Paydarfar et al.\cite{Paydarfar2006}
in their recent study recorded firing periods
in the range between 7 and 16 $\textrm{ms}$.
The overall dynamics of Figs. \ref{taka12}
and \ref{taka12-fit} agrees very well,
including the location and depth of the local minima.
We verified that the form of Fig. \ref{taka12-fit}
was unchanged for pulse widths between 0 and $1\textrm{ms}$
after dividing the current amplitude
by $\int_0^{T_i} I(t)dt$.



\begin{figure}[th]
\includegraphics[width=0.39\textwidth]{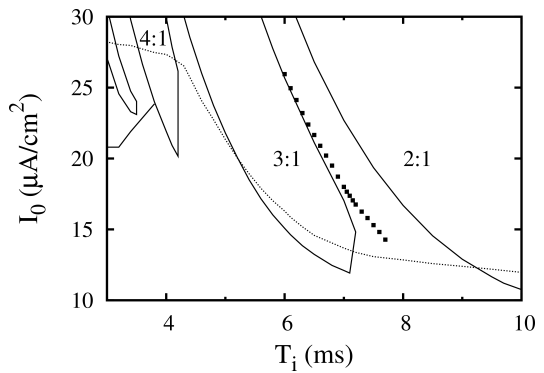}
\caption{The bifurcation diagram in the $T_i$-$I_0$ plane, showing
the main mode-locked states in the model without noise.
The unmarked intrusion in the upper left corner
is the 5:1 state. The bottom part of the figure
is occupied by the silent state.
In the firing part of the diagram there are two solutions
below the dotted line. Here the limit cycle coexists
with the fixed point.
Full squares show the location of the minima of the firing rate.
The borders of states below $T_i=4.5\textrm{ms}$
are shown in an approximate form. The detailed picture
is less regular and somewhat more complex.
}
\label{respdiag2}
\end{figure}

Fig. \ref{respdiag2} shows the response diagram
in the high-frequency limit. The dotted line
separates the monostable firing solution
from the silent state and bistable areas where the limit
cycle coexists with a fixed point solution.
Boundaries of bistability were determined using a continuation
method starting from a region with a single solution.

\begin{table}
\caption{\label{table}Frequency of occurrence
of the six lowest modes at the minimum
of the firing rate. The upper row is based
on Fig. 13e from the experimental data
of Takahashi et al.\cite{Takahashi1990}.
The bottom row is the result of calculations,
assuming $T_i=7\textrm{ms}$
and $I_0=18\mu \textrm{A/cm}^2$ (see Fig. \ref{modes}).}
\begin{ruledtabular}
\begin{tabular}{cccccc}
mode&&&&&\\
2&3&4&5&6&7\\
\hline
0&0.66&0.04&0.12&0.09&0.04\\
0.002&0.74&0.07&0.11&0.04&0.02
\end{tabular}
\end{ruledtabular}
\end{table}

\begin{figure}[th]
\includegraphics[width=0.38\textwidth]{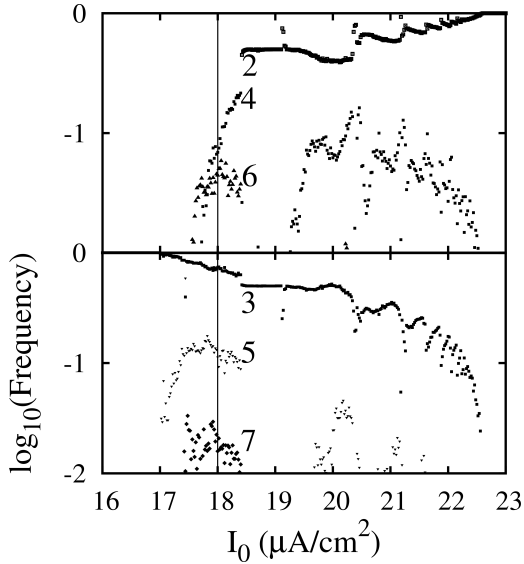}
\caption{The relative frequency of occurrence
of low-order even and odd modes for the parameter set
of Fig. \ref{taka12-fit}
The vertical line marks
the position of the minimum of the firing rate.
}
\label{modes}
\end{figure}

\begin{figure}[th]
\includegraphics[width=0.38\textwidth]{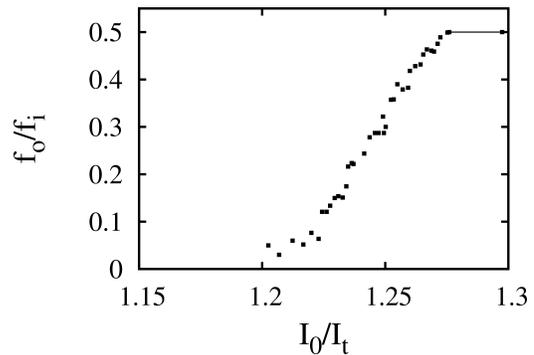}
\caption{The linear relation of the firing rate
vs. the stimulus amplitude above
the multimodal transition point at $T_i=4 \textrm{ms}$.
These are experimental results of Takahashi et al.\cite{Takahashi1990}.
}
\label{taka14}
\end{figure}

\begin{figure}[th]
\includegraphics[width=0.38\textwidth]{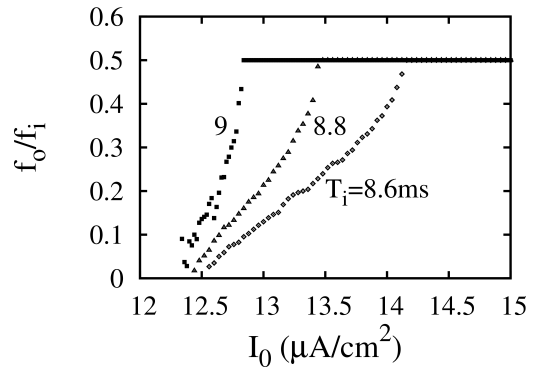}
\caption{Calculated average firing rate vs. stimulus amplitude
above the multimodal transition for three values of
$T_i$. The current pulse width is $0.6 \textrm{ms}$.
}
\label{taka14-fit}
\end{figure}

The experimental local maximum
on the plateau $f_o/f_i=0.4$ is due to the state
$\textrm{10100}$, where modes 2:1 and 3:1 alternate.
The other local maximum at $f_o/f_i=0.429$
with tendency to lock
into the $\textrm{(10)}^2100$
was also reproduced.
Fig. \ref{modes} shows the relative frequency of participation
of the most common modes on a logarithmic scale.
Higher-order modes appear more frequently
near the minimum of the firing rate.
Experimental and calculated ISI histograms are compared
in Table \ref{table}.
The overall agreement
is quite remarkable. Also the calculated evolution
of individual modes as a function of $I_0$
is close to measured values. In experiment
the probability of appearance of mode 4:1 between
$I_0/I_t=1.2$ and 1.3 remains in the range $0.06$ to $0.08$,
which agrees well with Fig. \ref{modes}
for $I_0$ between $18\mu\textrm{A/cm}^2$ and $22\mu\textrm{A/cm}^2$.
The published experimental runs\cite{Takahashi1990} contain 80 to 100
output spikes for selected data points.
On the basis of this data set we can conclude that the frequency
of participation of the low order modes is approximately
reproduced in simulations.
Above the multimodal transition
the experimental firing rate near the threshold rises linearly with
the stimulus amplitude, see Fig. \ref{taka14}.
The dependence of $f_o/f_i$ vs. of $I_0$
is well reproduced in Fig. \ref{taka14-fit},
except in the vicinity of the 2:1 plateau,
where an addition of a small amount of noise
would improve the fit.



We now consider the model with a Gaussian white noise:
\begin{equation}
\label{HHnoise}
C\frac{dV}{dt}=-I_{Na}-I_K-I_L+I_{app}+C \xi(t),
\end{equation}
where
$<\xi_i(t)>=0$, 
$<\xi(t)\xi(t^\prime)> = 2D\delta(t-t^\prime)$,
and $D$ is expressed in $\textrm{mV}^2/\textrm{ms}$.
The HH equations are integrated using the second-order stochastic
Runge-Kutta algorithm\cite{Honeycutt1992a}.
The simulations are carried out with the time step of $0.01 \textrm{ms}$
and are run for 400 s, discarding the initial 40 s.

\begin{figure}[th]
\includegraphics[width=0.38\textwidth]{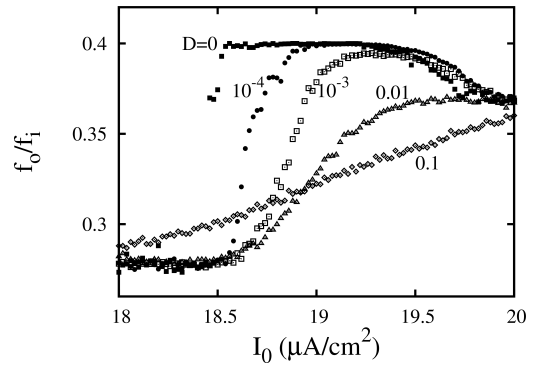}
\caption{Sensitivity of the $f_o/f_i=0.4$ plateau
from Fig. \ref{taka12-fit} to noise.
}
\label{plateau}
\end{figure}

There is a tendency to assume that biological systems,
including neurons, should always be treated as noisy systems.
While the neuron is sensitive to noise
it is not obvious that single neuron dynamics should
always include stochastic terms.
Fig. \ref{plateau} shows the quick disappearance
of the $f_o/f_i=0.4$ plateau in Fig. \ref{taka12-fit}
with increasing noise. Comparing with the experimental
data in Fig. \ref{taka12} we conclude that
calculations reproduce experimental data only
for $D<10^{-4}$.
Certainly more experiments
are needed to understand the role of noise in neurons.

\begin{figure}[th]
\includegraphics[width=0.38\textwidth]{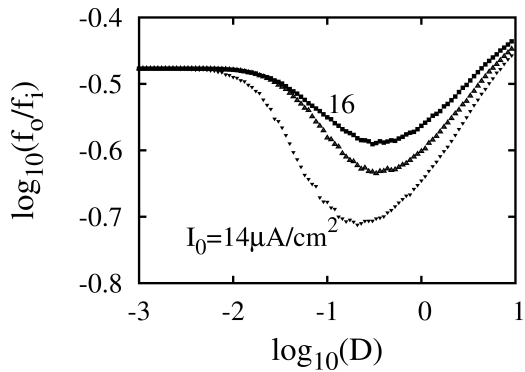}
\caption{The firing rate vs. the noise intensity.
The middle curve was obtained for $I_0=15\mu \textrm{A/cm}^2$.
Here $T_i=7\textrm{ms}$. At $D=0$ all three curves
start in the 3:1 mode.
}
\label{f-T=7}
\end{figure}

Fig. \ref{f-T=7} presents the firing rate as a function
of $D$ for three parameter sets from the $3:1$ plateau
of Fig. \ref{taka12-fit}.
For small noise $f_o/f_i$ drops quickly below
$1/3$ over an entire plateau,
with the biggest drop near the edges.
This behavior should be contrasted
with the resonant regime where the central part
of each plateau maintains phase locking over much
larger range of noise intensities
and $D\sim 1$ is needed to lower the firing rate of an entire
plateau below the $D=0$ value\cite{Borkowski2010}.
Another difference is
the direction of frequency changes at the plateau
edges. In the resonant state the frequency below (above)
the plateau midpoint is lowered (increased), respectively\cite{Borkowski2010}.
In the antiresonant limit the entire plateau is unstable
to even a small noise which slows down the system
considerably.


\begin{figure}[th]
\includegraphics[width=0.38\textwidth]{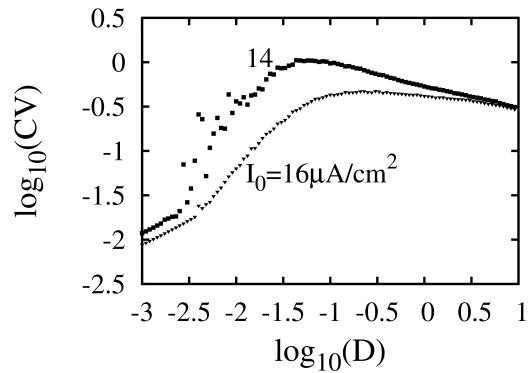}
\caption{The maximum of the coefficient of variation
as a function of the noise intensity is a property
of the stochastic coherence antiresonance.
The maximum of CV and the minimum of the firing rate occur
at different noise levels.
The irregularity
of the $I_0=14 \mu \textrm{A/cm}^2$ curve is
a consequence of proximity to the excitation threshold.
}
\label{cv-T=7}
\end{figure}

CV as a function of $D$
has a maximum for the same parameter set,
see Fig. \ref{cv-T=7}.
The increased variability is associated
with increased participation of higher-order modes
and may be called a \textit{stochastic coherence
antiresonance}.
A maximum of CV was found earlier
in a leaky integrate-and-fire model with an absolute
refractory period for suprathreshold base current\cite{Lindner2002}.
A small local maximum of CV at intermediate noise level
was also found by Luccioli et al.\cite{Luccioli2006}
in a HH model driven by a dc current in a bistable regime,
where the neuron was stimulated
by a large number of stochastic inhibitory and excitatory
postsynaptic potentials.
It was pointed out that the stochastic antiresonance
may exist in regions of bistability\cite{Gutkin2009},
when the stable limit cycle coexists with other attractors.
This typically occurs in the vicinity of a bifurcation
when the value of the bifurcation parameter slightly
exceeds the critical value.
In the HH model near the multimodal transition
there are many competing
limit cycles. Noise enhances trajectory
switching and may even stop the firing entirely.
A decrease of the firing rate and an increase of incoherence
may occur along much of the excitation threshold,
where the deterministic
system is bistable or responds irregularly\cite{Borkowski2010}.

In conclusion, numerical solutions
of the deterministic HH equations
show that the minimum of the firing rate
observed by Takahashi et al.\cite{Takahashi1990}
is due to the multimodal transition\cite{Borkowski2009}.
The statistics of the experimental
spike trains confirm that below the transition only odd
modes remain. Even modes are present at the minimum of $f_o/f_i$,
in agreement with theoretical calculations\cite{Borkowski2009}.
The calculated frequencies of occurrence
of the most common modes are close to experimental values.
Also the location of the minimum of $f_o/f_i$
in the vicinity of the 3:1 state is consistent with the simulations.
The linear rise of the output frequency as a function
of the stimulus strength above the multimodal
transition was also confirmed.
The excitation threshold in the antiresonant
limit is higher by about a factor of two compared
to the resonant regime.
The rise of threshold for frequencies
of current pulses exceeding the resonant frequency
was observed experimentally by Kaplan et al.\cite{Kaplan1996}.
Further support for the significance
of the parity of the modes comes from the experiment
of Paydarfar et al.\cite{Paydarfar2006}, who found
that the quiescent periods between highly regular bursts
were always equal to even multiples of the resonant period.
An ISI histogram with odd modes was obtained
by Racicot and Longtin\cite{Racicot1997} in a chaotically forced
FitzHugh-Nagumo (FHN) model. FHN equations
are often used as a substitute for the full HH model.
It would therefore be useful to investigate whether
the main features of the odd-all ISI transition
are reproduced in the FHN model with
a deterministic and stochastic drive.

Perturbing the system with
noise changes significantly the $f$ vs. $I_0$ dependence.
The local minima of this curve disappear already for
$D \simeq 10^{-3}$.
In the regime below the multimodal transition
the 3:1 plateau disappears rapidly for very small noise.
The firing rate has a minimum and CV
has a maximum as a function of the noise intensity.
These predictions are expected to be valid for short
stimuli of different shapes and can be tested
experimentally.
The multimodal transition and the accompanying
stochastic antiresonance are important
both for the understanding of excitable systems
and for potential neurological applications.
Spike annihilation by deterministic signals\cite{GuttmanLewisRinzel1980}
is studied in the context of deep brain stimulation\cite{Tass2002,Calitoiu2007},
a therapeutic technique, in which synchronicity
of certain parts of the brain is reduced by brief current pulses.




\end{document}